\documentclass[fleqn,twoside]{article}
\usepackage{espcrc2}

\usepackage{graphicx}
\usepackage{psfig}

\newcommand{\simlt}{\lower.5ex\hbox{$\; \buildrel < \over \sim \;$}}
\newcommand{\simgt}{\lower.5ex\hbox{$\; \buildrel > \over \sim \;$}}

\title{The faint HI environment of galaxies: a laboratory for
testing galaxy formation}

\author{Edvige Corbelli\address{INAF-Osservatorio 
Astrofisico di Arcetri, Firenze, Italy, edvige@arcetri.astro.it}}

\begin{document}

\begin{abstract}
The SKA is a unique instrument to open a window on many aspects of galaxy
formation and evolution which can be examined in our Local Universe.
Here I will focus on the outermost regions of galaxies which can be
observed with sufficient sensitivity and resolution at 21-cm as to enlighten
the interplay between galaxies and the intergalactic gas, the competing
race between the local dark matter gravity and the external ionizing radiation
field. Tracing the gas distribution out to large galactocentric radii
will be complementary to QSO's Ly$\alpha$ absorption studies for understanding 
the evolution of the dark and visible matter of the Universe.
\end{abstract}


\maketitle

\section{Introduction}

The faint gaseous envelopes of galaxies are important sites for
studying the processes which are at the heart of galaxy formation 
and evolution. They connect the bright active star forming
regions to their larger cosmic environment, made of gas, dark matter
and photons, through which galaxies are forming. 
The absence of strong star formation activity in these regions
makes them suitable for disentangling the cosmological variables
from local processes. Sensitive surveys of the 
outer regions of galaxies are needed in order to derive from them  
unique information, such as the interplay  
between galaxies and the intergalactic medium, 
the size and structure of dark matter halos, and the relevance of
collisional and photoionization processes for the evolution of 
cosmic structures. 
Possible local disturbances, such as those related to the proximity 
of a central star forming region or to close encounters between galaxies, 
can be fully understood and modeled only through a wide survey of galaxy 
morphologies and environments.

In the following Sections I will outline in more detail 
the need for observing the 21-cm emission
of the gas in the outer regions of galaxies down to column densities 
$\sim 10^{17}$~cm$^{-2}$ with the high sensitivity and resolution 
that SKA will provide.
Current deep 21-cm surveys with single dish telescopes are sensitive 
to column densities $N_{HI}\simeq 1-5\times 10^{18}$~cm$^{-2}$ but 
the large 
beam width limits the information one can derive on the kinematics and
distribution of the neutral gas (Minchin et al. 2003).
High resolution imaging with beam sizes $\leq$ 1$^{\prime}$ has
usually been done for column densities higher than  
$2\times 10^{19}$~cm$^{-2}$ (e.g. Hunter \& Wilcots 2002,
Wilcots \& Hunter 2002).  
The actual number of galaxies mapped with spatial resolution
of order of 1~kpc and a sensitivity in the range 1-5$\times 10 
^{18}$~cm$^{-2}$ is very limited since this requires  a large amount of 
telescope integration time or the galaxy must be located in our close 
neighborhood (confusion with emission of the Milky way halo is of major
concern in this case).

\section{21-cm emission maps and QSO Ly-alpha absorption lines:  
different probes of similar structures in the evolving Universe}

\subsection{Intervening absorption systems in QSO spectra}

Current progress in the theoretical
framework have been made to understand high redshift optical
observations which gave indirect evidence of the presence of
faint condensations along the line of sight to QSOs.
These condensations have been modeled as filaments connecting the
IGM to the high density contrast central regions, which is where
star formation takes place. Gas condensations may flow in and out the 
galaxy potential well and direct imaging of the 21-cm emission 
will hopefully enlighten the radial components of the velocity field 
of the galaxy low density environments (see Braun, this Volume, for 
addressing this SKA science case).
However there is no compelling evidence 
of these structures yet and it is also not well understood how they 
connect to the brighter central regions. 

Although 
Lyman-$\alpha$ absorption line studies 
are well suited for tracing the column density distribution of the HI gas
at various cosmological epochs for column densities 
$N_{HI}< 10^{17}$~cm$^{-2}$. Statistics for higher column densities 
are more difficult.
This is because Lyman-$\alpha$ absorption lines are not suitable for
a determination of $N_{HI}$ when $10^{17}<N_{HI}<10^{20}$~cm$^{-2}$  
and lines are in the flat part of the 
curve of growth. Gas clouds in this column density regime show 
optical depths $\tau > 1$ to the quasar ionizing radiation and 
will make a break in the quasar spectrum at the frequency
corresponding to the Lyman edge (Lyman limit systems). Only for
$\tau \sim 1$~ can N$_{HI}$ be measured
accurately.  Larger $\tau$ values imply that Lyman limit systems
can be detected but N$_{HI}$ cannot be measured.
Therefore the HI column density distribution in this regime is poorly 
known, not only because of the limits on today's 21-cm emission surveys,
but also because of the undetermined value of $N_{HI}$ in
intervening QSO absorption systems.

\subsection{Ionization processes}

It is of great importance to have detailed 
information on the HI column density distribution for
$10^{17}<N_{HI}<10^{20}$~cm$^{-2}$. This lies not only on the 
relevance of knowing the baryonic and dark matter
boundaries of galactic type structures, but also on the
fact that this regime connects the mostly neutral part of the
gas distribution to the highly ionized side of it.
Therefore knowing the details of the gas distribution in
this regime will open a window on the ionization processes  
acting in the outskirts of galaxies. Important ingredients for 
galaxy formation, such as the UV background radiation
field, the escaping fraction of UV photons from star
forming disks, and the presence of collisional ionization from violent
relaxation will be enlightened.
This information is also essential for deriving the total
gas column density distribution and hence the gas
baryonic content of our Local Universe.
Theoretical models and the two high resolution observations
of the HI distribution along one side of the major axis of
M33 and NGC3198 have shown that in outer disks $N_{HI}$ drops
by one order of magnitude in less than one beam size
(about 1 and 3~kpc respectively for the two observations)
as the column density approaches the face on value of
$2\times 10^{19}$~cm$^{-2}$ (Corbelli \& Salpeter 1993, Maloney 
1993). This drop has
been interpreted in terms of a sharp HI/HII transition due to
photoionization from UV extragalactic background radiation.
Modeling a sharp edge in detail for deriving the
extragalactic radiation field intensity requires a knowledge
of the possible contamination from nearby UV sources,
of the radial scale length of the HI gas distribution
where the gas is mostly neutral, and a knowledge of the
dark matter density near the edge (the latter 
because the ionization-recombination balance
depends also on the gas volume density which is a function 
of the gravitational potential).
This explains the need for observing a large sample of
objects to disentangle all the variables.
A new statistical approach for deriving some 
information on the ratio between the ionization field intensity
and the gravity in intervening QSO absorbers 
has been developed by Bandiera $\&$ Corbelli (2001).
However, it is only by inferring the 
properties of the hosting dark matter halo through accurate 
21-cm maps of the velocity field in outer regions that the
radiation field intensity  can be recovered. Figure 1
shows this together with the need for N$_{HI}$ data close
to the HI photoionization
threshold for drawing definite conclusions on the ionization
processes and on the total density of baryons in our Universe.
The next generation of H-$\alpha$ detectors will certainly
complement the above research (e.g. Tufte et al. 2002).

\begin{figure}
\psfig{figure=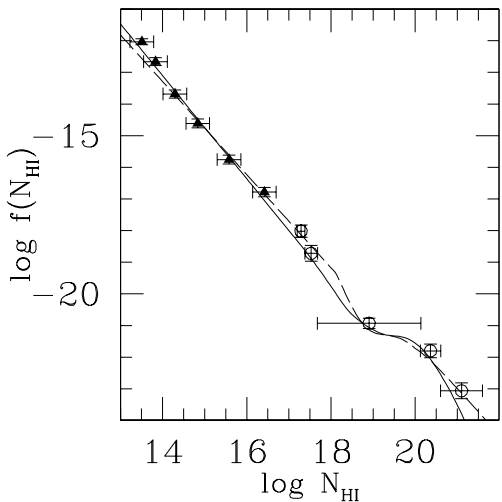,width=3in}
\caption{Data and two best fitting models for the N$_{HI}$ distribution 
function in our Local Universe. Filled triangles are from the compilation of
Ly$\alpha$ forest data by Weymann et al. (1998), open circles are 
from the Lyman-limit and damped Ly$\alpha$ systems catalog
of Corbelli et al. (2001). Both the continuous and dashed line functions
fit the actual data quite well but they correspond to two different
ionization models and to two different total gas densities (see
Corbelli \& Bandiera (2002) for details). This figure shows the
need for HI data for N$_{HI}>10^{17}$~cm$^{-2}$ to constrain the models.}
\end{figure}

\subsection{Dark matter halos}

Today's sensitivities and resolutions, together with the uncertainties
on the stellar mass to light ratio, do not allow us to test
the validity of dark matter models which predict a cusp in the
center of dark matter halos. The consistency of these models 
with the actual data requires generally very low values of the halo
concentration (e.g. de Blok et al. 2001, 
van den Bosch $\&$ Swaters 2001, Corbelli 2003). This in turn implies 
that rotation curves should be sampled with higher resolution in
the inner regions but especially they should be extended to larger 
radii in order to reach regions where the dynamical contribution
of the stellar disk is negligible compared to that of the halo. 
Many galaxies 
today still show a smooth rising or flat rotation curve  
at the outermost sampled radius.
By extending maps of the outer regions by two orders of magnitude down in 
HI column density with respect to today's measurements we will 
be able to  detect the stronger 
decline of the dark matter density with radius.
The most promising cases for detecting the decline of the rotation curve
will be those galaxies with the highest value of dark matter concentration.

\section{First experiment: beyond the star forming regions}

\subsection{Goals}

Through observations of the outskirts of galaxies, described in this 
Section, one should be
able to derive important information on the dark matter, on
the extragalactic ionizing radiation and other ionizing processes
in the Local Universe. In particular one can establish: 

--If there is a flattening in the  total $N_{HI}$ distribution for
  $N_{HI}<10^{20}$~cm$^{-2}$ and the HI column density where it possibly 
  occurs. This flattening should be present if photoionization
  processes are relevant in the outskirts of galaxies.

--If there are ionization edges in the HI distribution of single  
  galaxies and if their properties are universal or depend on local  
  and environmental conditions. 

--How does the amplitude and the smoothness of rotation 
  vary with galactocentric radius well outside the region where
  the stellar disk dynamics is dominant.

--If there is any evidence of gas circulation from or towards 
  the brighter star forming regions.

--The outer disks morphology: thermal phases, clouds mass,
  spiral arms, flaring.

\subsection{Observing strategies}

We need to observe a large sample of galaxies
in order to disentangle the extragalactic
background ionizing radiation from other sources of ionization
(UV photons escaping from nearby star forming regions,
turbulent mixing layers from collisional ionized hot gas...)
and to avoid the possible degeneracy with an undefined dark matter
density distribution.
The ideal sample will comprise galaxies of similar masses but
with different levels of star formation activity as well as galaxies
with different estimates of the dark matter density in the
outer regions (accessible with today's sensitivities) and in
different environments.
In order to define the structural properties of dark matter halos,
galaxies with a low ratio of visible to dark matter and
with extended HI disks need to be sample. Also important is
to extend the HI maps of those galaxies which show high values
of the dark matter concentration since for those galaxies the
probability of detecting the boundary of the dark halo is
higher.

In order to build up an N$_{HI}$ distribution function our 
observation should sample the whole range of
HI mass with a well known HI mass function.
Galaxies should be sampled with a spatial resolution comparable with
the HI scale length. If the ionization edge happens to be
present, then the scale length of the HI distribution  is
about 1~kpc
and observations of outer regions for this large sample  should
be carried out with a limiting N$_{HI}$ sensitivity of 
$10^{18}$ cm$^{-2}$.

The observations should be carried out with a channel width 
lower than the thermal speed of the medium and
which would also be sufficient for deriving the kinematic
information. 
Taking 2 km~s$^{-1}$ spectral channel resolution for a typical line
width of 30 km~s$^{-1}$, a SKA spatial resolution of 30~arcsec 
requires an integration time of about 4~hours for a 3-$\sigma$
detection of 10$^{18}$~cm$^{-2}$ column densities, given the
actual design specifications. Target galaxies will be
located in the nearby 10~Mpc Universe but the total large SKA 
bandwidth will sample for free a much deeper volume in $z$.
Therefore this experiment can be done in conjunction with searches
for very low surface brightness objects.

\section{Second experiment: from galaxies to the IGM}

\subsection{Goals}

In this second experiment one would like to observe the 21-cm 
emission which originates from  HI gas at column densities as
low as $10^{17}$~cm$^{-2}$. In this regime  the following features
should be detectable:

--Where do disks end i.e. how disks connect with IGM filaments and 
  if there is evidence of material falling into the galaxies
  from the intergalactic medium. 

--The slope of the N$_{HI}$ distribution function for 
  N$_{HI}\sim 10^{17}$~cm$^{-2}$. Here the ionization
  fraction is high and it does not vary so rapidly with the total 
  gas column density as for higher N$_{HI}$ values.
  This, together with QSO absorption 
  data for lower HI column densities and with N$_{HI}$ data from
  the first experiment, will allow us to define unambiguously the
  HI-Htot relation and therefore the total gas content of our
  Universe (see Figure 1). 
  If the dark matter-gas scaling relation holds, one 
  can derive the matter density and check the compatibility of the 
  ionization models with the adopted cosmology. 

\subsection{Observing Strategies}

From the large survey done in the first experiment, 
observers pick up the most representative candidates for 
carrying out the deeper integration.  As one goes towards HI
column densities lower than 10$^{18}$~cm$^{-2}$ (where the 
HI/HII transition is taking 
place), the steeper slope of the N$_{HI}$ distribution function
ensures that the sky area filled with such a low density gas will
increase as N$_{HI}$ decreases. Therefore one doesn't need to survey
large sky areas as for the first experiment. The likely spatially
fragmented HI distribution which will characterize this regime  
implies however that a few areas should be sampled for averaging
out local perturbations and orientation effects and reducing the 
errors in the N$_{HI}$ distribution. 

With a beam size as large as 100~arcsec observers should be able 
to sample the gas distribution with a sensitivity (at the 3$\sigma$ 
level) which is 10$^{17}$~cm$^{-2}$ for 21 hours of integration (assuming a
50 km~s$^{-1}$ signal width, and 10~km~s$^{-1}$ spectral resolution).
This experiment can be done in conjunction with filament searches 
in the local Universe.

\end{document}